\documentclass{article}

\usepackage{graphicx}

\begin{document}

\title{On surface tension for compact stars}
\author{R. Sharma\thanks{206526115@ukzn.ac.za}~ \& S. D.
Maharaj\thanks{maharaj@ukzn.ac.za}\\
Astrophysics and Cosmology Research Unit\\ School of Mathematical Sciences,
University of KwaZulu-Natal\\ Private Bag X54001, Durban 4000, South Africa.}

\maketitle

\date{}

\begin{abstract}
In an earlier treatment it was demonstrated that general relativity
gives higher values of surface tension in strange stars with quark matter
than neutron stars.We generate the modified Tolman-Oppenheimer-Volkoff
equation to incorporate anisotropic matter and use this to show that
pressure anisotropy provides for a wide range of behaviour in the surface
tension than is the case with isotropic pressures. In particular it is
possible that anisotropy drastically decreases the value of the surface tension.
\end{abstract}

$Keywords:$ Relativity-pulsars-equation of state

\section{Introduction}
Stars which are more compact than neutron stars, at present,
have become a subject of considerable interest as they provide us
natural laboratories for testing QCD. Over the last couple of decades,
various models have been proposed to explain the compactness and
properties of some of the observed compact objects. Pioneering works
in this field have put forward new concepts of compact matter, namely
strange stars (Witten 1984, Farhi \& Jaffe 1984) and boson stars (Kaup
1968, Ruffini \& Bonazzola 1969, Colpi {\em et al} 1986). Due to the high
matter densities within such stars one expects pressure to be anisotropic
in general, i.e., in the interior of such stars the radial pressure and
tangential pressure are different. An anisotropic energy momentum is a topic
 which is often ignored in the calculations of compact stars. However,
 since the pioneering work of Bowers \& Liang (1974) there has been extensive
 research in the study of anisotropic relativistic matter in general relativity.
 The analysis of static spherically symmetric anisotropic fluid spheres is
 important in relativistic astrophysics.  Ruderman (1972) showed that nuclear
 matter may be anisotropic in the high density ranges of order
 10$^{15}$~gm~cm$^{-3}$ where nuclear interactions have to be treated
 relativistically. Anisotropy in compact objects may occur due to the
 existence of a solid core or the presence of type 3A superfluid
 (Kippenhahm \&  Weigert 1990), phase transition (Sokolov 1980),
 pion condensation (Sawyer 1972), slow rotation (Herrera \& Santos 1997),
 mixture of two gases (Letelier 1980) or strong magnetic fields (Weber 1999).
 Also objects made up of self-interacting  scalar particles known as boson
 stars are naturally anisotropic in their configurations. Anisotropic models
 for compact self gravitating objects have been studied by Herrera \& Santos
 (1997), Rao {\em et al} (2000), Corchero (2001), Mak \& Harko (2003),
 Ivanov (2002), Dev \& Gleiser (2003),
 Hern$\acute{a}$ndez, \& N$\acute{u}$$\widetilde{n}$ez (2004),
 Chaisi \& Maharaj (2005), and many others. Anisotropic models for
 compact objects have been shown to achieve high red-shift values
 (Bowers \& Liang 1974, Herrera \& Santos 1997, Ivanov 2002, Mak \&
  Harko 2003), and they are stable (Herrera \& Santos 1997, Dev \&
  Gleiser 2003). In this article, we show that pressure anisotropy
  may also effect the surface tension of compact stars. We believe that
  this aspect has not been considered yet in the context of anisotropic
  stellar models.

\section{Surface Tension of strange stars}
In a recent paper by Bagchi {\em et al} (2005), it has been shown that
objects composed of $u$, $d$ and $s$ quarks popularly known as `strange stars'
give higher values of surface tension than neutron stars, a necessary criterion
for the existence of stable strange stars in the Universe. This calculation is
based on equations of state (EOS) for strange matter formulated by
Dey {\em et al} (1998). In an approximated linearized form, the EOS
may be written as (Zdunik 2000, Gondek-Rosi\'{n}ska {\em et al} 2000)
\begin{equation}
p = a(\rho - \rho_{b}) \label{eq1}
\end{equation}
where $\rho$ is the energy density, $\rho_{b}$ is the density at the
surface, $p$ is the isotropic pressure, and $a$ is a parameter related
to the velocity of sound ($a=dp/d\rho$).

To calculate the surface tension, one assumes that the star is a huge spherical
ball composed of strange matter which is self-bound and non-rotating. The excess
pressure on the surface of the star can be expressed as
\begin{equation}
|\Delta p|_{r=R} = \frac{2 S}{R}\label{eq2}
\end{equation}
where $S$ is the surface tension of the star and $R$ is the radius of curvature.
At the surface
\begin{equation}
|\Delta p|_{r=R} = r_{n}\frac{dp}{dr}|_{r=R} \label{eq3}
\end{equation}
where, $r_{n}$ is the radius of the quark particle given by
$r_{n} = (1/\pi n)^{1/3}$ where $n$ is the baryon number density.
 As strange stars are very compact, a relativistic treatment is
 necessary to find their configurations and other physical parameters.
 Thus for a given EOS, one uses the Tolman-Oppenheimer-Volkoff (TOV)
 equation (Oppenheimer \& Volkoff 1939)
\begin{equation}
\frac{dp}{dr} = -\frac{G(\rho + p)\left[\frac{m(r)}{c^2 r}
+ \frac{4\pi r^2 p}{c^4}\right]}{c^2 r \left(1-
\frac{2Gm(r)}{ r}\right)} \label{eq4}
\end{equation}
to find the surface tension of the star, making use of
equations (\ref{eq2}) and (\ref{eq3}). This method helps
to yield higher values of surface tension as compared to
neutron stars including the possible explanation for the
existence of strange stars in the Universe and other
related phenomena like delayed $\gamma$-ray bursts (Bagchi {\em et al} 2005).

However, at very high densities, anisotropy may be
significant in such stars which may contribute to the
surface tension. If we assume that pressure within such a star
is anisotropic in general then the TOV equation (\ref{eq4}) gets
 modified yielding different results as obtained by Bagchi {\em et al}
 (2005). In the following sections, we derive the modified TOV equation
 with anisotropic pressure and perform some numerical calculations to
 show the effects of pressure anisotropy on the surface tension of compact stars.

\section{Anisotropic TOV equation}
We first formulate the modified TOV equation with
anisotropic pressure. We assume the line element for
a static spherical object in the standard form
\begin{equation}
ds^{2} = -e^{\gamma (r)}c^2dt^{2} + e^{\mu (r)}dr^{2}
+ r^{2}(d\theta^2 + \sin^2\theta d\phi^2)\label{eq5}
\end{equation}
where $\gamma(r)$ and $\mu(r)$ are the two unknown metric functions.
Without any loss of generality, the energy momentum tensor for an
anisotropic star may be written as
\begin{equation}
T_{ij}= (\rho c^2 + p_{r})u_{i}u_{j}+p_{r}g_{ij}+(p_{r}-p_{\perp})n_{i}n_{j}\label{eq6}
\end{equation}
where $u_{i}$ is the fluid four-velocity, $n_{i}$ is a radially
directed unit space-like vector. We assume that  $p_{r}\neq p_{\perp}$
and $p_{\perp}- p_{r} = \Delta$ gives the measure of pressure anisotropy in this model.

The Einstein's field equations are then given by
\begin{eqnarray}
\frac{8\pi G}{c^4}\rho &=& \frac{\left(1-e^{-\mu}\right)}{r^2}
+\frac{\mu'e^{-\mu}}{r} \label{eq7} \\
\frac{8\pi G}{c^4}p_{r} &=& \frac{\gamma'e^{-\mu}}{r}-\frac{\left(1-
e^{-\mu}\right)}{r^2} \label{eq8}\\
\frac{8\pi G}{c^4}p_{\perp} &=& \frac{e^{-\mu}}{4}\left(2\gamma''
+{\gamma'}^2-\gamma'\mu'+\frac{2\gamma'}{r}-
\frac{2\mu'}{r}\right)\label{eq9}
\end{eqnarray}
where primes denote differentiation with respect to the radial
coordinate $r$. Equations (\ref{eq7})-(\ref{eq9}) may be combined
together to yield
\begin{equation}
(\rho + p_{r})\gamma' + 2p_{r}'+\frac{4}{r}(p_{r} - p_{\perp}) = 0 \label{eq10}
\end{equation}
which is a conservation equation.

If we write the metric function $\mu$ in terms of mass function $m(r)$ as
\begin{equation}
e^{-\mu} = 1 - \frac{G m(r)}{c^2 r} \label{eq11}
\end{equation}
then equation (\ref{eq10}) becomes
\begin{equation}
\frac{dp_{r}}{dr} = - \left(\rho + p_{r}\right)\frac{\left(\frac{G m(r)}{c^2 r}
 + \frac{4\pi G r^2 p_{r}}{c^4}\right)}{r \left(1-\frac{2G m(r)}{c^2 r}\right)}
  + \frac{2}{r}(p_{\perp}-p_{r}) \label{eq12}.
\end{equation}
Equation (\ref{eq12}) is the the modified TOV equation in the presence of
pressure anisotropy. For a given central density $\rho_{c}$ or central
pressure $p_{r}^c$ and anisotropic parameter $\Delta$, equation (\ref{eq12})
 may be integrated to find the mass $M = m(R)$ and radius $R$ of the star provided
 the EOS $p_{r} = p_{r}(\rho)$ is known. Local anisotropy thus effects the
 geometry of the star.

At the surface of the star $r=R$, the radial pressure $p_{r}$ vanishes. However,
the tangential pressure $p_{\perp}$ is not necessarily zero at the surface. The
two pressure profiles within the star should satisfy the following
conditions: $p_{r} > 0$ and $p_{\perp} > 0$. The maximum value of
the anisotropic parameter $\Delta$ vis a vis the tangential pressure
 $p_{\perp}$ is constrained by the physical requirement that the radial
 pressure gradient $dp_{r}/dr$ should be negative in the stellar interior;
 other physical requirements may, however, put a more stringent restriction
 on the values of $\Delta$.  Thus for finite values of $p_{\perp}$ at the
 boundary $\Delta(r=R) = p_{\perp}^b$, equation (\ref{eq12}) becomes
\begin{equation}
\frac{dp_{r}}{dr}|_{r=R} = - \frac{\rho_{b}\frac{G M}{c^2 R}}{R \left(1-
\frac{2G M}{c^2 R}\right)} + \frac{2p_{\perp}^b}{R} \label{eq13}.
\end{equation}
If $p_{\perp}^b$ is not negligible at the boundary, equation (\ref{eq13}) shows
that it is possible to get different sets of values of surface tension as
obtained by Bagchi {\em et al} (2005) for isotropic matter. Thus it is
possible to generate a wide range of behaviour in the surface tension for
anisotropic matter than is the case for isotropic pressures.

\section{Numerical results}
To get an estimate of the effects of pressure anisotropy on the surface
tension, we consider the strange matter EOS given by equation (\ref{eq1}).
We consider two particular cases as discussed by Gondek-Rosi\'{n}ska {\em et al} (2000):

(i) EOS SS1: where, $a=0.463$, $\rho_{b} = 1.15\times 10^{15}$~gm~cm$^{-3}$,
$\rho_{c} = 4.68\times 10^{15}$~ gm~cm$^{-3}$, $n(r=R)=0.725$~fm$^{-3}$,
$n(r=0)=2.35$~fm$^{-3}$, $M=1.435~M_{\odot}$, $R=7.07$~km.

(ii) EOS SS2: where, $a=0.455$, $\rho_{b} = 1.33\times 10^{15}$~gm~cm$^{-3}$,
$\rho_{c} = 5.5\times 10^{15}$~ gm~cm$^{-3}$, $n(r=R)=0.805$~fm$^{-3}$,
$n(r=0)=2.638$~fm$^{-3}$, $M=1.323~M_{\odot}$, and $R=6.55$~km.

\begin{figure}
\centering
\includegraphics[scale=0.8]{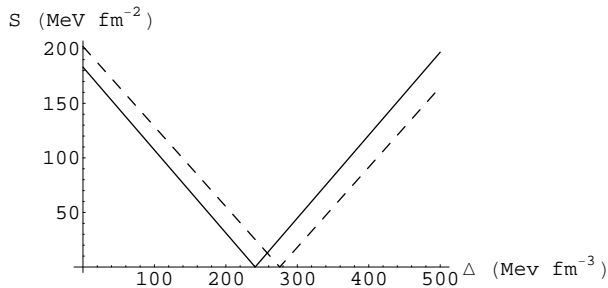}
\caption{\label{fig1} Surface tension $S$ plotted
against $\Delta$. The solid line is for EOS SS1 and
the dotted line is for EOS SS2.}
\end{figure}

Numerical calculations show that for a given mass and radius, if we
gradually introduce anisotropy, the absolute value of the surface
tension decreases as can be seen from figure~\ref{fig1}. For example,
it is observed that even if we consider a tangential pressure of
$100$~MeV~fm$^{-3}$ at the surface, the surface tension decreases
drastically. It is to be noted here that the anisotropy parameter
should be so chosen that all the regularity conditions (Delgaty \& Lake 1998)
are satisfied. Thus, although in figure~\ref{fig1}, the surface tension
increases beyond a certain value of the anisotropic parameter, we ignore
this region as the  radial pressure gradient becomes positive in this
region. The results are given in Table 1.
\begin{table*}
\centering
\caption{Anisotropic effect on the surface tension of strange stars.}
{\begin{tabular}{|l|c|c|c|c||r|@{}} \hline
EOS & $r_{n}$~(fm) & \multicolumn{2}{|c|}{$\frac{dp_{r}}{dr}|_{r=R}$~(MeV~fm$^{-3}$~km$^{-1})$} &  \multicolumn{2}{|c|}{$S$~(MeV~fm$^{-2})$}\\ \hline
     & &  $ p_{\perp} = 0$  & $p_{\perp}=100$~(MeV~fm$^{-3})$ &  $p_{\perp}=0$ & $p_{\perp} = 100$~(MeV~fm$^{-3})$ \\ \hline
SS1 & 0.76 & 68.18 & 39.90 & 183.19 & 107.19 \\
SS2 & 0.73 & 84.09 & 53.56 & 202.14 & 128.75 \\ \hline
\end{tabular}}
\end{table*}

\section{Discussions}
We have shown that anisotropy plays an important role in the
calculation of surface tension of compact stars. The origin of
such anisotropies within compact objects may be different for
different objects. We may, however, ask whether it is necessary
at all to consider anisotropic effects on the surface tension of
strange stars. The answer is affirmative since one possibility for
the origin of anisotropies within strange stars could be the presence
of charged particles at the surface. It has recently been reported
that in strange stars, the electric field could be as high as
$10^{19}~$eV/cm (Usov 2004), which indicates the possibility of
a large charge distribution within such objects. Therefore we need
to consider the effect of charge while deriving the the gross features
of such stars. It can be shown that in the presence of charge, the TOV
equation is modified to
\begin{equation}
\frac{dp}{dr} = - \left(\rho + p\right)\frac{\left(\frac{G m(r)}{c^2 r}
+ \frac{4\pi G r^2 p}{c^4}\right)}{r \left(1-\frac{2G m(r)}{c^2 r}\right)}
+ \frac{Q(r)}{4\pi r^4}\frac{dQ(r)}{dr} \label{eq14},
\end{equation}
where, $Q(r)$ is the total charge confined within a sphere of radius $r$.
Note that Einstein-Maxwell system is always anisotropic which is often
treated as an isotropic system of field equations for mathematical
simplicity (see for example, Ray {\em et al} 2004). Also recent works
(Schmitt 2005) suggest that a natural mechanism to explain the strong
pulsar kicks in neutron stars could be the existence of asymmetric phases
in quark matter.

It is to be noted here that, for simplicity,  we ignored the effect
of rotation in the present work although pulsars are magnetized
rotators and a strong magnetic field ($\sim 10^{12}~$G) is observed
at the surface of such stars. Pulsars known as magnetars may even
have magnetic field as strong as $\sim 10^{14-15}~$G. Though we do
not have an established theory for the microscopic origin of such a
strong magnetic field, it is agreed that Ferro-magnetization may
occur in the high density quark matter which, in turn, may modify
the EOS for strange matter. The derivation and the form of the
modified EOS in the presence of strong magnetic field or
superfluidity (responsible for anisotropy) is a complex issue and a
more detailed analysis is required to see the effect of the modified
EOS on the overall configuration vis a vis surface tension of
compact stars.

To conclude, without going into the microscopic details of a star,
it can be shown that surface tension is affected in the presence of anisotropy.
For the very existence of strange stars in our Universe a crucial condition put
forward was a large value of $S$ by Alcock \& Olinto (1989) which according to
Bagchi {\em et al} (2005) can be achieved by a general relativistic treatment
of strange stars. However, in this article we have shown that a wide range of
values of $S$ are possible if we consider anisotropy in the energy momentum
tensor; an issue ignored in the previous calculation (Bagchi {\em et al}
2005). Therefore, on the basis of surface tension for compact stars, perhaps
no conclusive remarks at this moment can be made on the possible existence of
strange stars. There could be, however, some other means to justify the
existence of such stars which will be taken up elsewhere.

\section*{Acknowledgments}
RS acknowledges the financial support (grant no. SFP2005070600007) from
the National Research Foundation (NRF), South Africa. SDM acknowledges that
this work is based upon research supported by the South African Research Chair
Initiative of the Department of Science and Technology and the National
Research Foundation.

\end{document}